# Performance and Energy-Aware Bi-objective Tasks Scheduling for Cloud Data Centers

Huned Materwala[1,2] and Leila Ismail[1,2,*]

[1]Intelligent Distributed Computing and Systems Research Laboratory, Department of Computer Science and Software Engineering, College of Information Technology, United Arab Emirates University, Al Ain, Abu Dhabi, 15551, United Arab Emirates
[2]National Water and Energy Center, United Arab Emirates University, Al Ain, Abu Dhabi, United Arab Emirates

## Abstract

Cloud computing enables remote execution of users' tasks. The pervasive adoption of cloud computing in smart cities' services and applications requires timely execution of tasks adhering to Quality of Services (QoS). However, the increasing use of computing servers exacerbates the issues of high energy consumption, operating costs, and environmental pollution. Maximizing the performance and minimizing the energy in a cloud data center is challenging. In this paper, we propose a performance and energy optimization bi-objective algorithm to tradeoff the contradicting performance and energy objectives. An evolutionary algorithm-based multi-objective optimization is for the first time proposed using system performance counters. The performance of the proposed model is evaluated using a realistic cloud dataset in a cloud computing environment. Our experimental results achieve higher performance and lower energy consumption compared to a state-of-the-art algorithm.

**ACM – Class**: I.2.1; I.2.8

**Keywords:** Autonomous agents; Cloud computing; Energy-efficiency; Evolutionary algorithm; Genetic algorithm; Quality of service; Multi-objective optimization; Performance

## 1. Introduction

Cloud computing [1] has become a very promising paradigm for both consumers and service providers allowing convenient, on-demand network access to a shared pool of configurable computing resources. With the advancement in technological paradigms such as the Internet of Things (IoT) and Big data analytics for smart cities' applications, data center traffic is exploding with the rapid growth of cloud applications. It is predicted that the global cloud data center traffic will increase from 6 zettabytes (ZB) in 2016 to reach 19.5 ZB by the year 2021 [2]. Furthermore, with the current COVID-19 pandemic situation, all the essential services such as healthcare, work, food, and education have become online. These services heavily rely on the cloud computing paradigm. Consequently, cloud computing infrastructure must maintain large-scale data centers, consisting of thousands of computing nodes that consume a large amount of electrical power.

---
[*] Correspondence: Leila Ismail (email: leila@uaeu.ac.ae)

It is estimated that the data centers will become the world's largest energy consumers globally, with an increase from 3% of total energy consumption in 2017 to 4.5% in 2025 [3]. The data center energy cost increases by 100% every 5 years [4]. High energy consumption not only incurs a high cost but also harms the environment. It is predicted that by 2025 the data centers will emit nearly 3.5% of carbon emission globally [5]. According to a report by Natural Resources Defense Council (NRDC), it is expected that data centers will emit nearly 100 million metric tons of carbon pollution per year [6]. Consequently, it becomes crucial to address this issue of cloud energy consumption.

Several works in the literature have proposed energy-efficient tasks' scheduling algorithms using evolutionary algorithm in the cloud computing environment [7]–[12]. As tasks' scheduling in the cloud is an NP-hard problem, the evolutionary algorithm, such as genetic [13], is well suited for task optimization problems due to its characteristics of parallel and efficient global search. However, the tasks' performance, i.e., Quality of Service (QoS) should be considered while minimizing energy consumption. Very few works in the literature focus on multi-objective performance and energy-aware tasks scheduling in the cloud using evolutionary algorithm [8]–[12]. However, none of these works considers the tasks' resource utilization in terms of system performance metrics, i.e., CPU, memory, disk, and network, while computing the energy consumption. This is crucial considering the dynamic nature of the tasks submitted to the cloud.

In this paper, we develop an intelligent autonomous agent for performance and energy-aware bi-objective tasks' scheduling in a cloud data center based on the evolutionary algorithm. We consider the task's execution time as a measure of performance. The tasks' scheduling is modeled as a bi-objective optimization problem to minimize tasks' execution time and energy consumption. We use the Locally Corrected Multiple Linear Regression (LC-MLR) [14] power consumption model, which is based on CPU, memory, disk, and network utilization, for the prediction of the computing server's power consumption. The predicted power is then used to compute the server's energy consumption. The performance of the proposed model is evaluated using a realistic cloud dataset in terms of energy consumption and execution time. This is in a cloud data center simulated using the CloudSim 3.0.3 [15], a software tool for cloud computing simulation. The performance of the proposed model is compared with the genetic algorithm-based task scheduling model in the literature that uses a power model based on CPU and memory utilization values [8].

The rest of the paper proceeds as follows. Section 2 provides an overview of the related work. The cloud system model is presented in Section 3. Section 4 describes the optimization problem and its formulation using evolutionary algorithm. The experiments and the performance evaluation are presented in Section 5. Section 6 concludes our work.

## 2. Related Work

Several works in the literature have proposed the use of the evolutionary genetic algorithm for energy-efficient multi-objective tasks' scheduling in a cloud computing environment [8]–[12]. However, [11] and [12] do not mention the power model used for the computation of energy consumption. A hardware-based power model using the computing server's voltage and frequency is considered by [10]. However, the hardware-based power model often requires physical sensors for monitoring the hardware resources. This leads to high hardware cost and sensors' energy consumption when the sensors are attached to thousands of servers in a cloud data center [16]. A software-based power model consisting of system performance metrics such as CPU, memory, disk, and/or network resources is used by [9] and [8]. However, the power model used by [9] is based only on CPU utilization, and the one used by [8] is based on CPU and memory utilization values. To the best of our knowledge, none of the works on performance and energy-optimized cloud tasks' scheduling based on evolutionary genetic algorithm use an energy consumption formulation based on system performance counters. In this work, we propose an evolutionary algorithm-based intelligent agent for task scheduling in cloud computing while minimizing the task's execution time and energy consumption. The energy consumption in the proposed bi-objective optimization method considers system performance counters. We compare the performance of our proposed model with the genetic algorithm-based bi-objective optimization model in the literature that uses power model based on CPU and memory utilization values [8].

## 3. Architecture Overview

The cloud computing architecture consists of 'v' heterogeneous virtual machines (VMs) that operate on 'p' heterogeneous physical machines (PMs) as shown in Figure 1. The set of VMs in represented as V = {$VM_1$, $VM_2$, …, $VM_v$} and the set of PMs is represented as P = {$PM_1$, $PM_2$, …, $PM_p$}. The cloud users' tasks are submitted to the cloud broker which implements an intelligent agent that schedules the tasks on a VM such that the energy consumption and task execution time are the minima. The task analyzer monitors and records the resources and service requirements of the tasks submitted by the cloud users. The resources' requirements of a task include the CPU, memory, disk, and network utilization values, while the service requirement involves the performance metrics such as task deadline and execution time. Based on the task's requirements in terms of CPU, memory, disk, and network, the agent calculates the execution time and energy consumption on each VM. Therefore, the agent communicates with the VM manager which is responsible to monitor the resource utilization of running VMs. It reads the current energy consumptions of the VMs which are maintained by the energy consumption monitor of the cloud.

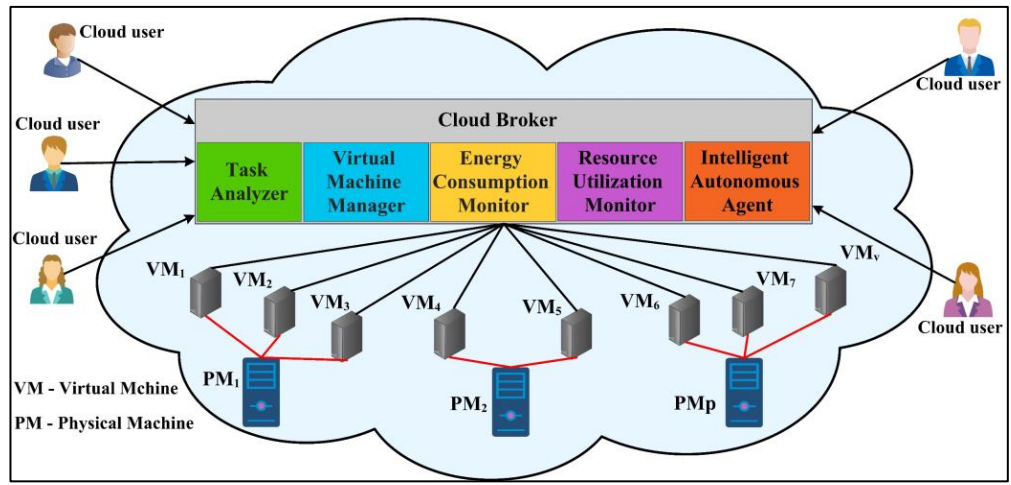

*Figure 1: An intelligent autonomous agent-based cloud architecture overview.*

A virtual machine in our architecture is characterized by Processing Speed (PS) in terms of Million Instructions Per Second (MIPS). A task $t_j$ is characterized by its length, $L(t_j)$, in Million Instructions (MI), percentage CPU utilization, memory utilization in MB, disk utilization in bytes read/write per second, and network utilization in bytes transferred per second. The execution time $ET_{j,i}$ of task $t_j$ on $VM_i$ is computed using Equation 1.

$$ET_{j,i} = \frac{L(t_j)}{PS_i} \times n_i \tag{1}$$

where $PS_i$ is the processing speed of $VM_i$ and $n_i$ is the number of tasks running simultaneously on $VM_i$ including $t_j$.

The energy consumption of task $t_j$ is computed by multiplying the task's power consumption $P_{CPU_{j,i},mem_{j,i},disk_{j,i},net_{j,i}}$ and execution time on $VM_i$ as stated in Equation 2.

$$EC_{j,i} = P_{CPU_{j,i},mem_{j,i},disk_{j,i},net_{j,i}} \times ET_{j,i} \tag{2}$$

The power consumption of executing a task on a computing server is predicted using a power model. We use the Locally Corrected Multiple Linear Regression (LC-MLR) power model as stated in Equation 3. LC-MLR is selected in this paper because it is found to be accurate in a cloud computing environment [14].

$$P_{CPU_{j,i},mem_{j,i},disk_{j,i},net_{j,i}} = P'_{CPU_{j,i},mem_{j,i},disk_{j,i},net_{j,i}} + \emptyset_{CPU_{j,i},mem_{j,i},disk_{j,i},net_{j,i}} \tag{3}$$

where $P'_{CPU_{j,i},mem_{j,i},disk_{j,i},net_{j,i}}$ is the predicted power consumption using the MLR regression model as stated in Equation 4 and $\emptyset_{CPU_{j,i},mem_{j,i},disk_{j,i},net_{j,i}}$ is the error correction term as stated in Equation 5.

$$P'_{CPU_{j,i},mem_{j,i},disk_{j,i},net_{j,i}} = \alpha + (\beta_1 \times CPU_{j,i}) + (\beta_2 \times mem_{j,i}) + (\beta_3 \times disk_{j,i}) + (\beta_4 \times net_{j,i}) \tag{4}$$

$$\begin{aligned}
\emptyset_{CPU_{j,i},mem_{j,i},disk_{j,i},net_{j,i}} &= e_{CPU_i^k,mem_i^l,disk_i^m,net_i^n} \\
&+ \left[\frac{\left(e_{CPU_i^{k+1},mem_i^l,disk_i^m,net_i^n} - e_{CPU_i^k,mem_i^l,disk_i^m,net_i^n}\right)(CPU_{j,i} - CPU_i^k)}{(CPU_i^{k+1} - CPU_i^k)}\right] \\
&+ \left[\frac{\left(e_{CPU_i^k,mem_i^{l+1},disk_i^m,net_i^n} - e_{CPU_i^k,mem_i^l,disk_i^m,net_i^n}\right)(mem_{j,i} - mem_i^l)}{(mem_i^{l+1} - mem_i^l)}\right] \\
&+ \left[\frac{\left(e_{CPU_i^k,mem_i^l,disk_i^{m+1},net_i^n} - CPU_i^k,mem_i^l,disk_i^m,net_i^n\right)(disk_{j,i} - disk_i^m)}{(disk_i^{m+1} - disk_i^m)}\right] \\
&+ \left[\frac{\left(e_{CPU_i^k,mem_i^l,disk_i^m,net_i^{n+1}} - e_{CPU_i^k,mem_i^l,disk_i^m,net_i^n}\right)(net_{j,i} - net_i^n)}{(net_i^{n+1} - net_i^n)}\right]
\end{aligned} \tag{5}$$

where $\alpha, \beta_1, \beta_2, \beta_3$ and $\beta_4$ are the regression coefficients, $CPU_{j,i} \in [CPU_i^{k+1}, CPU_i^k]$, $mem_{j,i} \in [mem_i^{l+1}, mem_i^l]$, $disk_{j,i} \in [disk_i^{m+1}, disk_i^m]$, and $net_{j,i} \in [net_i^{n+1}, net_i^n]$. $e_{CPU_i^k,mem_i^l,disk_i^m,net_i^n}$, $e_{CPU_i^{k+1},mem_i^l,disk_i^m,net_i^n}$, $e_{CPU_i^k,mem_i^{l+1},disk_i^m,net_i^n}$, $e_{CPU_i^k,mem_i^l,disk_i^{m+1},net_i^n}$, and $e_{CPU_i^k,mem_i^l,disk_i^m,net_i^{n+1}}$ are the errors calculated as the difference between the actual and the predicted power consumption values obtained from the MLR model.

The energy consumption of a task $t_j$ on $VM_i$ can be then calculated using Equation 6, based on the energy function proposed in [17] that considers the increase in the energy consumption of the ongoing tasks on a VM due to the increase in their execution time while calculating the energy consumption of a new task on that VM.

$$EC_{j,i} = \begin{cases} P_{CPU_{j,i},mem_{j,i},disk_{j,i},net_{j,i}} \times ET_{j,i}, & VM_i \text{ is idle} \\ P_{CPU_{j,i}+CPU_{j-1,i},mem_{j,i}+mem_{j-1,i},disk_{j,i}+disk_{j-1,i},net_{j,i}+net_{j-1,i}} \times ET'_{j,i}, & ET'_{j,i} < NET_{j-1,i} \\ \left(P_{CPU_{j,i}+CPU_{j-1,i},mem_{j,i}+mem_{j-1,i},disk_{j,i}+disk_{j-1,i},net_{j,i}+net_{j-1,i}} \times \tau_0\right) + \left(P_{CPU_{j,i},mem_{j,i},disk_{j,i},net_{j,i}} \times \tau_1\right), & ET'_{j,i} > NET_{j-1,i} \end{cases} \quad (6)$$

where $NET_{j-1,i}$ is the new execution time of the task $t_{j-1}$ that was ongoing on $VM_i$ while task $t_j$ is scheduled on $VM_i$. The new execution time is the increment in the execution time of $t_{j-1}$ as the processing speed of $VM_i$ is distributed among $t_{j-1}$ and $t_j$. $ET'_{j,i}$ is the execution time of $t_j$ when running in parallel with $t_{j-1}$. $\tau_0$ is the time when the task $t_j$ is executed in parallel with the task $t_{j-1}$ and $\tau_1$ is the time when the task $t_j$ is executed alone.

## 4. Performance and Energy-Aware Bi-objective Tasks Scheduling

Let us consider a set of tasks, T = {t$_1$, t$_2$, …, t$_m$} that needs to be scheduled on a set of virtual machines, V. The scheduling of the tasks on the VMs is represented using a matrix S(m x v). For instance, Sji = 1 indicates that the task tj is scheduled on VM$_i$ for execution. The bi-objective optimization problem is to schedule tasks in a cloud computing environment in a way that the execution time and the energy consumption are the minima. These objectives are represented using a weighted sum cost function as stated in Equation 7.

$$CF_{j,i} = [\alpha \times ET_{j,i}^{norm}] + [(1-\alpha) \times EC_{j,i}^{norm}] \quad (7)$$

where $\alpha$ and $(1-\alpha)$ are the weights for the execution time and the energy consumption objectives respectively such that $0 \leq \alpha \leq 1$.

The bi-objective tasks scheduling optimization problem can be now formulated as follows:

Objective:
$$\forall j \in T \; Minimize(CF_{j,i}), i = \{1, 2, 3, \ldots, v\} \quad (8)$$

Constraints:
$$\sum_{j=1}^{m} S_{j,i} = 1, i = \{1, 2, 3, \ldots, v\} \quad (9)$$

$$\forall i \in v \sum_{j=1}^{m} CPU_{j,i} < CPU'_{j,i} \quad (10)$$

where Equation 8 shows the optimization objective, i.e., minimizing the cost function. Equations 9 and 10 represent the constraints. Equation 9 states that each task should be executed only on one VM and Equation 10 indicates that the total utilization of a VM should be always less than a threshold utilization to avoid performance degradation.

### 4.1 Bi-objective Evolutionary Algorithm using Intelligent Autonomous Agent

The task scheduling optimization problem in a cloud computing environment can be designed as an autonomous agent system where the agent schedules the tasks on the VMs to minimize the objective function. The task analyzer, the VM manager, the power consumption monitor, and the resource utilization monitor components in the cloud broker represent the sensors of the system environment and the mapping of tasks to the VMs depict the actuator output. The agent's system environment for task scheduling is fully observable, stochastic, sequential, dynamic, discrete, and single-agent. The intelligent autonomous agent for task scheduling can be classified as a utility-based agent [18] as shown in Figure 2. This is because the tasks' scheduling problem involves contradicting optimization objectives with a trade-off between energy consumption and execution time. Evolutionary genetic algorithm [13] is a search-based heuristic. The main components of the evolutionary algorithm are as follows:

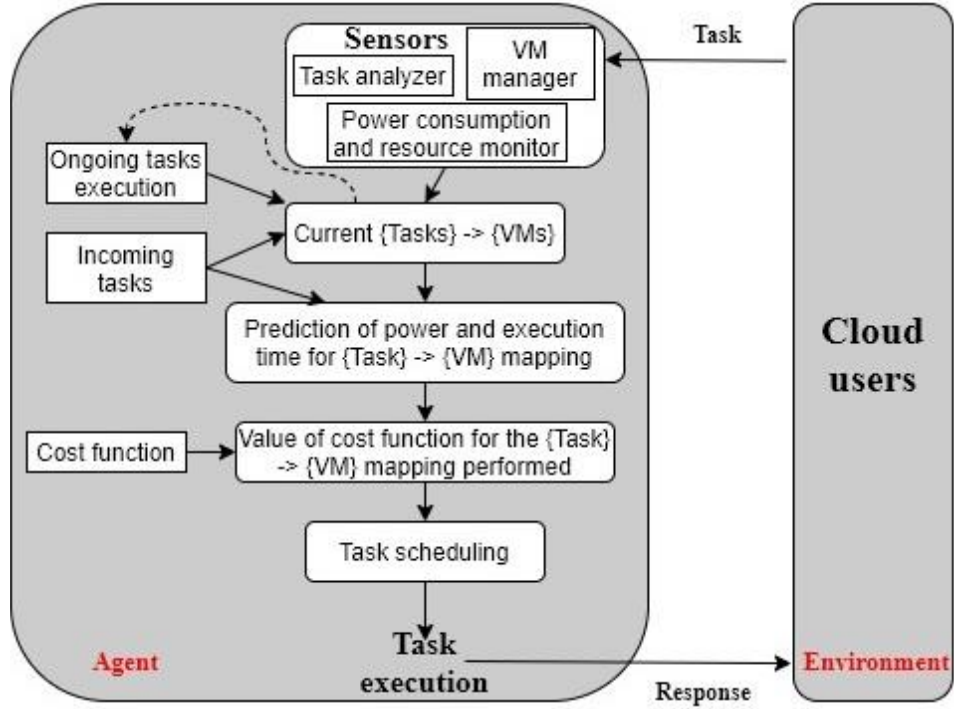

*Figure 2: Utility-based intelligent autonomous agent for tasks scheduling.*

- *Initial tasks-VMs mapping (population):* The mapping of tasks to the VMs is the initial population in cloud tasks' scheduling. Each solution in the population is represented as a chromosome. The chromosome for tasks scheduling problem can be considered as the mapping of tasks to VMs.
- *Fitness function:* The inverse of the cost function for task scheduling that minimizes the energy consumption and the execution time (Equation 7) is the fitness function for the problem under study.
- *Crossover:* Crossover operation is achieved by selecting two parent population and then creating a new mapping by alternating some or all the genes of the parents. Each element of the chromosome is known as a gene.
- *Mutation:* It is the operator that produces offspring by tweaking genes of a single chromosome.

In this paper, we use an energy-efficient task scheduling algorithm Modified Worst Fit Decreasing (MWFD) for the selection of the initial population. MWFD is chosen for the selection of the initial population due to its optimal performance compared to other energy-aware task scheduling algorithms [19]. This reduces the time to obtain a global solution. In MWFD, each task is assigned to a VM where the increase in power consumption after scheduling the task is the maximum. Algorithm 1 shows the pseudocode for population initialization. Algorithm 2 shows the pseudocode for bi-objective optimization using evolutionary algorithm.

| Algorithm 1: Population initialization using MWFD |
| --- |
| 1. **Input**: TaskList, VMList **Output**: Scheduling of Tasks |
| 2. TaskList.sortByCpuUtilization_decreasing() |

|  |  |
|---|---|
| 3. | **foreach** Task in TaskList **do** |
| 4. |    maxPower = Double.MIN_VALUE |
| 5. |    allocated VM = null |
| 6. |    **foreach** VM in VMList **do** |
| 7. |      **if** VM has enough resources for Task **then** |
| 8. |        powerAfterAllocation = Calculate power using Equation 7 |
| 9. |        powerDiff = VM.getPower() - powerAfterAllocation |
| 10. |        **if** powerDiff >maxPower **then** |
| 11. |           allocatedVM = VM |
| 12. |           maxPower = powerDiff |
| 13. |    **if** allocatedVM ≠ null **then** |
| 14. |      allocation.add(Task, allocatedVM) |
| 15. | **Return** allocation |

**Algorithm 2: Performance and energy-aware bi-objective tasks' scheduling using evolutionary algorithm**

1. **Input**: TaskList, VMList **Output**: Scheduling of Tasks
2. Generate the initial tasks and VMs mapping using **Algorithm 1**
3. **while** (non-termination condition) **do**
4.    SelectFitTasksVMsMapping //select initial tasks-VMs mapping
5.    Perform_crossover_NewTasksVMsMapping //create new scheduled mapping
6.    Perform_mutation
7.    **foreach** newMapping **do** //check for each new scheduled tasks
8.      **if** Fitness.newMapping<Fitness.Mapping **then** //check the fitness value of the new mapping
9.        add.Mapping(newMapping) //if new mapping is more efficient than the previous one then add the new mapping to the list
10.      replace.currentMapping(feasiblenewMapping)
11. **Return** allocation //return the new tasks and VMs mapping for the scheduled tasks

## 5. Performance Evaluation

### 5.1 Experimental Environment

To evaluate the performance of the proposed model in a cloud computing environment with a large number of hosts and VMs, we simulate a cloud data center using CloudSim 3.0.3 simulation software [15]. We create a homogenous data center by using six different host types (Table 1) and five different VM types (Table 2). Servers 1 and 2 from the host types are part of our Intelligent Distributed Computing and Systems (INDUCE) research laboratory at the College of Information technology of the United Arab Emirates University. The specifications of servers 3-6 are taken from the SPEC Power benchmark suite [20] in a way that they belong to the same family as the ones present in the laboratory, but with different architectures and capabilities.

*Table 1: List of host types used in the experiments.*

| | |
|---|---|
| Server 1 | Sun Fire Intel_Xeon CPU core of 2.80 GHz, Dual-core, with 512 KB of cache and 4 GB of memory for each core, CPU voltage rating 1.5 V, OS version CentOS 6.8(i686) |
| Server 2 | Sun Fire X4100 with AMD_Operaton252 CPU of 2.59 GHz, dual CPU, single-core, with 1MB of cache and 2GB of memory for each core, CPU voltage rating of 3.3-2.9 V, OS version Red Hat Enterprise Linux Server release 7.3 (Mapio) |
| Server 3 | Dell Inc. PowerEdge R260 with Intel Xeon E5-2670 CPU core of 2.6 GHz CPU, 8 core, with 2MB cache, 4GB 2Rx8 PC3L10600E-9 ECC memory and 1 x 100GB SATA SSD disk drive [21] |
| Server 4 | SGI Rackable C2112-4G10 with AMD Opteron 6276 CPU core of 2.30 GHz, 16 cores, 4GB 2Rx8 PCL-10600R memory and 1 x 120 GB 2.5" SSD SATA disk drive [22] |
| Server 5 | Hewlett Packard Enterprise ProLiant DL360 Gen9 with Intel Xeon E5-2699 v3 CPU core of 2.30 GHz, 18 core, with 45MB L3 Cache, 8 GB 2Rx8 PC4-2133P memory and 1 x 400GB SSD SATA disk drive [23] |
| Server 6 | Acer Incorporated Acer AR585 F1 with AMD Opteron 6238 CPU core of 2.60 GHz, 12-core, with 16 MB L3 cache, 4GB 2Rx8 PC3L-10600E memory and 1 x 500GB SATA2 7200 RPM 3.5" HDD disk drive [24] |

*Table 2: List of VM types used in the experiments.*

| | VM_type 1 | VM_type 2 | VM_type 3 | VM_type 4 | VM_type 5 |
|---|---|---|---|---|---|
| {MIPS, RAM (MB)} | {2200, 870} | {1800, 1740} | {2000, 870} | {1500, 1740} | {1750, 613} |

## 5.2 Experiments

To evaluate the performance of our proposed evolutionary algorithm scheduler using LC-MLR (GA_LC-MLR), we first simulate the data center with an appropriate number of hosts and VMs. We create 800 hosts with each of the host types equally distributed. We then create the VMs with each of the four VM types equally distributed. We generate tasks with the random length between 6000-12500 MI at an interval of 3 seconds each. For the CPU utilization of the tasks, we use the real-life workload data traces from CoMon project, a monitoring infrastructure for PlanetLab [25]. The workload consists of the CPU utilization values collected from more than 1000 VMs from servers located at 500 different places across the globe. We use the data of 3 March 2011 for the experiments. For the memory, disk, and network utilization values we randomly generate memory sizes, read/write bytes per second, and data transferred per second, respectively. We compare the performance of GA_LC-MLR with the genetic algorithm in the literature that uses a linear power model based on CPU and memory utilization (GA_LM) [8]. We measure the total energy consumption for the tasks' execution and mean execution time.

## 5.3 Experimental Results Analysis

Figure 3 shows the energy consumption of GA_LC-MLR and GA_LM. It shows that the energy consumption of GA_LC-MLR is low. This is because of two reasons. First, the proposed algorithm considers the impact of ongoing tasks running on a server while scheduling an incoming task on that server. Consequently, the algorithm selects a server where the increase in the energy consumption is the minimum considering the execution of the tasks when running alone and when running in parallel with the task to place. Second, the energy consumption in the proposed algorithm considers the CPU, memory, disk, and network resources utilization values whereas the GA_LM algorithm considers only the CPU and memory utilization values. Consequently, the proposed model predicts more accurately the power consumption of a server while scheduling a task compared to the GA_LM algorithm, leading to higher energy savings. Figure 4 shows the comparison of mean execution time. GA_LC-MLR takes 0.5 seconds less compared to GA_LM.

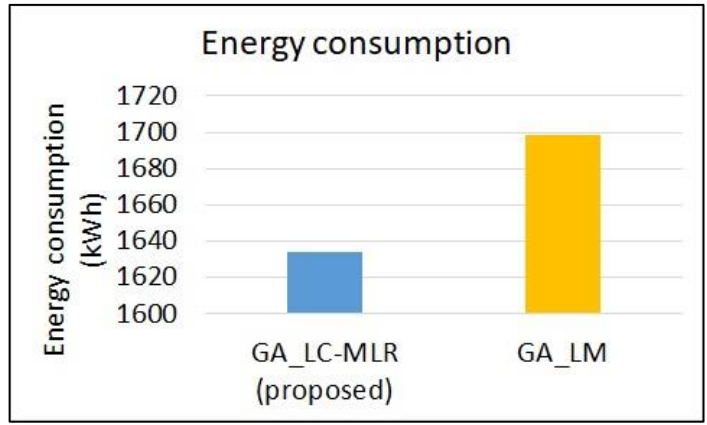

*Figure 3: Energy consumption using GA_LC-MLR and GA_LM.*

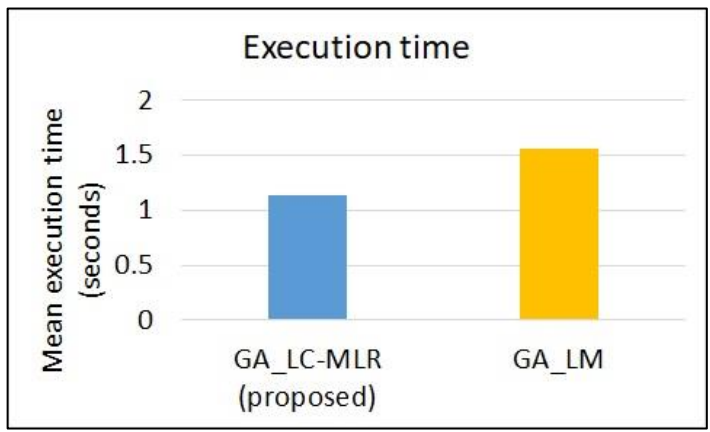

*Figure 4: Mean execution time using GA_LC-MLR and GA_LM.*

## 6. Conclusion

Cloud computing is an emerging technology, enabling companies to consume a computing resource as a utility rather than building and maintaining an in-house infrastructure. Due to the development of smart cities and worldwide pandemics, the use of cloud computing is increasing. This extensive utilization of cloud computing resources leads to the high energy consumption of underlying data centers. In addition to electricity consumption, environmental threats become considerable. In this work, we proposed an intelligent autonomous agent scheduler that schedules a user's task in a way that the energy consumption and execution time are minimized. The developed bi-objective optimization model is based on evolutionary algorithm. We evaluate the performance of the model in terms of energy consumption and execution time.

## Acknowledgements

This research was funded by the National Water and Energy Center of the United Arab Emirates University (Grant 31R215).